\documentstyle[12pt,a4,amstex]{article}
\newcommand\slv{v\kern-5pt\raise1pt\hbox{$\scriptstyle/$}\kern1pt}

\newcommand\be{\begin{equation}}
\newcommand\ee{\end{equation}}
\newcommand{\eps}{\varepsilon}
\def\bq{\begin{eqnarray}}
\def\eq{\end{eqnarray}}

\newlength{\dinwidth}                                                           
\newlength{\dinmargin}                                                          
\begin{document}

\thispagestyle{empty}
\begin{flushright}
NIKHEF-99-024\\
\end{flushright}
\vspace{0.5cm}
\begin{center}
{\Large \bf Massive gauge bosons from the conservation of topological winding numbers
}\\[.3cm]
\vspace{1.7cm}
$\mbox{{\sc \bf Stefan Weinzierl$^{1}$}}$
\begin{center} \em 
NIKHEF, P.O. Box 41882, NL - 1009 DB Amsterdam, The Netherlands \\
\end{center}\end{center}
\vspace{2cm}

\begin{abstract}\noindent
{We consider a $U(1) \times SU(2)$ gauge theory on the four-dimensional
manifold $S^1 \times S^3$.
If we make the assumption that only gauge transformations connected to the
identity are allowed, the winding numbers of $U(1)$ around $S^1$ and of
$SU(2)$ around $S^3$ become topological conserved quantities.
We derive the effective theory for non-trivial winding numbers if
all distances are small compared to the radii of the spheres.
In the non-abelian case the gauge bosons become massive.
}    
\end{abstract}
\vspace{3cm}

\vspace*{\fill}

\noindent 
$^1${\small email address : \verb/stefanw@nikhef.nl/}\\

\newpage

\section{Introduction}

The connection between topology and masses
has already a long history,
starting with the work of Schwinger \cite{Schwinger}, who showed that in two-dimensional
QED the photon acquires a mass through the polarization diagrams of (massless) fermion
loops.
In three dimensions massive gauge bosons can be obtained by adding a Chern-Simons term
to the action \cite{Deser}.
The discovery of instanton solutions \cite{Belavin} - \cite{Callan} 
revealed a non-trivial vacuum in four-dimensional
non-abelian gauge theories, and explained the mass of the $\eta$-meson.\\
\\
In this paper we consider a $U(1) \times SU(2)$ gauge theory on the four-dimensional
manifold $S^1 \times S^3$. For the metric we assume Euclidean signature.
We will further assume that we can savely neglect instanton effects.
It is obvious that this manifold lacks $SO(4)$--Lorentz symmetry, since
the coordinate along the circle $S^1$ is singled out.
However it is an interesting toy model for a $U(1) \times SU(2)$ gauge theory:
$S^1$ allows maps with non-trivial winding numbers into $U(1)$, as does
$S^3$ into $SU(2)$. In the context of gauge theories these maps correspond
to pure gauge configurations, and two configurations with different winding
numbers are considered to be equivalent, since they can be obtained from each other
by a gauge transformation. Of course this gauge transformation has to change
the winding number.\\
\\
On the other hand it is an interesting question to study the case in which we restrict
ourselves to gauge transformation which do not change the winding numbers.
If we make the assumption that only gauge transformations connected to the
identity are allowed, the winding numbers of $U(1)$ around $S^1$ and of
$SU(2)$ around $S^3$ become topological conserved quantities.\\
\\
In this paper we would like to make three assumption:
First, as already mentioned, we assume that we can neglect instanton effects.
Our second assumption is that only gauge transformations 
connected to the identity are allowed. \\
Requiring a non-trivial winding number, for example, of  $SU(2)$ around $S^3$ 
does not tell us anything
about the variation of the background field along the direction of $S^1$.
Our third assumption is therefore that the $SU(2)$-background field varies along 
$S^1$ randomly enough, such that we may replace it by an average over all $SU(2)$-rotated
configurations, which have the required winding number around $S^3$.\\
\\
We derive the effective theory for non-trivial winding numbers
if all distances are small compared to the radii of the spheres, but large
compared to the characterisitic length required by assumption 3.
We find that the $U(1)$-photon stays massless and that the $SU(2)$-gauge bosons develop
massive modes. We would like to point out that no additional scalar fields are involved.
Allthough in this paper we restrict ourselves to a pure gauge theory (e.g. we do not include 
fermions and we make no
attempt to explain a non-zero value for the Weinberg angle),
we would like to mention that J.A. Bagger, A.F. Falk and M. Swartz \cite{Bagger} have argued recently
that the present electroweak precision measurements are consistent with theories in which
there are no new particles below $3 \mbox{TeV}$.\\
\\
Mathematically we are considering secondary characteristic classes \cite{Chern}.\\
\\
This paper is organized as follows: 
In order to keep the discussion as simple as possible we first introduce 
in the next section a toy model, which relies only on the first two assumptions and for which the
$SU(2)$-background field stays constant along $S^1$.
In section 3 we derive in a general way the effective action in the presence of a background field.
In section 4 we apply these results to our toy model.
In section 5 we relax the unneccessary restriction, which is inherent in our specific toy model,
namely that the $SU(2)$-background field stays constant along $S^1$,
and replace it by
our assumption 3. We obtain a Lorentz-invariant theory with massive $SU(2)$-gauge bosons.
The conclusions are given in section 6.
The appendix contains some formulae for self-dual and anti-self-dual tensors.

\section{The toy model}

We embed $S^1 \times S^3$ in ${\Bbb R}^2 \times {\Bbb R}^4$ as follows
\bq
\label{embedding}
y_0 = r \cos \phi_0, & & y_1 = r \sin \phi_0, \nonumber \\
y_2 = r \cos \theta_2, & & y_3 = r \sin \theta_2 \cos \theta_3, \nonumber \\
y_4 = r \sin \theta_2 \sin \theta_3 \cos \phi_1, & & y_5 = r \sin \theta_2 \sin \theta_3 \sin \phi_1.
\eq
$\phi_0$ is the spherical coordinate for $S^1$, $\phi_1, \theta_2$ and $\theta_3$ are the
spherical coordinates for $S^3$. $\phi_0$ and $\phi_1$ take values in $[0,2\pi]$, whereas
$\theta_2$ and $\theta_3$ take values in $[0,\pi]$.
For simplicity we have assumed that the radii of the spheres are equal.
The metric tensor on $S^1 \times S^3$ is given by
\bq
g & = & r^2 d\phi_0^2 + r^2 \sin^2 \theta_2 \sin^2 \theta_3 d\phi_1^2
+ r^2 d\theta_2^2  + r^2 \sin^2 \theta_2 d\theta_3^2.
\eq
In the neighbourhood of $\phi_0=0, \phi_1=0, \theta_2=\pi/2, \theta_3 = \pi/2$ we introduce
a local coordinate system $X'=(x_0',x_1',x_2',x_3')$ by
\bq
x_0' = r \tan \phi_0, & & x_1' = r \tan \phi_1, \nonumber \\
x_2' = r \tan \left( \theta_2 - \frac{\pi}{2} \right), & & x_3' = r \tan \left( \theta_3 - \frac{\pi}{2} \right). 
\eq
Finally we relate the coordinate system $X'=(x_0',x_1',x_2',x_3')$ to a coordinate
system $X=(x_0,x_1,x_2,x_3)$ through a rotation:
\bq
x_\mu' & = & \Lambda_{\mu\nu} x_\nu,
\eq
where $\Lambda_{\mu\nu} \in SO(4)$.
It is obvious that the coordinate systems $X$ and $X'$ do not cover the entire manifold $S^1 \times  S^3$ but only
some chart $U$.
\\
Maps with non-trivial winding number from $S^1$ to $U(1)$ and from $S^3$ to $SU(2)$ are
given by
\bq
\tilde{B} & : & S^1 \rightarrow U(1), \nonumber \\
          & & \phi_0 \rightarrow \exp\left( i n \phi_0 \right), 
\eq
and by
\bq
B & : & S^3 \rightarrow SU(2), \nonumber \\
  & & (\phi_1,\theta_2,\theta_3) \rightarrow \frac{1}{r^m} \left( y_2 1 + i y_3 \sigma_1
+ i y_4 \sigma_2 + i y_5 \sigma_3 \right)^m, 
\eq
where $y_2$, ..., $y_5$ are given in eq. (\ref{embedding}),
$n$ and $m$ are the (integer) winding numbers and the $\sigma_a$ are the Pauli matrices.
We calculate $\tilde{B}^{-1} \partial_\mu \tilde{B}$ and
$B^{-1} \partial_\mu B$ in the coordinate system $X$:
\bq
\label{background}
\tilde{B}_\mu = \tilde{B}^{-1} \partial_\mu \tilde{B} & = &
 i \frac{n}{r} \Lambda_{0\mu} + O\left(\frac{1}{r^3}\right) \nonumber \\
B_\mu = B^{-1} \partial_\mu B & = & i \frac{m}{r}
\left( \sigma_1 \Lambda_{1\mu} + \sigma_2 \Lambda_{2\mu} + \sigma_3 \Lambda_{3\mu} \right)
+ O\left( \frac{1}{r^3} \right)
\eq
We now consider a pure Yang-Mills theory on $S^1 \times S^3$. The action is given
by
\bq
\label{Lagrangian}
\hat{S} = \int\limits_{S^1 \times S^3} d^4x \sqrt{|g|}
\hat{{\cal L}}_{YM}, & & \hat{{\cal L}}_{YM} = \frac{1}{2 g^2}  
\mbox{Tr} \; \hat{F}_{\mu\nu} \hat{F}^{\mu\nu}
\eq
with
\bq
\hat{F}_{\mu\nu} & = & \partial_\mu \hat{A}_\nu - \partial_\nu \hat{A}_\mu 
+ \left[ \hat{A}_\mu, \hat{A}_\nu \right], \nonumber \\
\hat{A}_\mu & = & T^a \hat{A}^a_\mu.
\eq
The $T^a$ are the hermitian generators of the gauge group. We take the
normalization to be
\bq
\mbox{Tr} \; T^a T^b & = & \frac{1}{2} \delta^{ab}.
\eq
For $SU(2)$ we have therefore $T^a = \frac{1}{2} \sigma^a$ and
\bq
\label{strucconst}
\left[ T^a, T^b \right] = i f^{abc} T^c, & & f^{abc} = \frac{1}{2} \eps_{abc}.
\eq
The explicit expressions for the $U(1)$- and the $SU(2)$-gauge potentials are
\bq
\label{background2}
\tilde{B}_\mu = \frac{2 i n}{r} \Lambda_{0\mu}, & &
B^a_\mu = \frac{2 i m}{r} \Lambda_{a\mu},
\eq
up to terms of order $1/r^3$.
We split the gauge field $\hat{A}_\mu$ into a fluctuating field $A_\mu$ 
and a background field $B_\mu$: 
\bq
\hat{A}_\mu & = & A_\mu + B_\mu
\eq
The latter one is given explicitly by
eq. (\ref{background2}).
Since the background field is pure gauge ($B_\mu = B^{-1} \partial_\mu B$) it has
vanishing curvature:
\bq
\partial_\mu B_\nu - \partial_\nu B_\mu + \left[ B_\mu, B_\nu \right] & = & 0
\eq
We will further assume that the fluctuating field $A_\mu$ vanishes outside a region
$V$ which is entirely contained in the local chart $U$.
In that case we may replace the integration over $S^1 \times S^3$ 
by the integration over $V$.
We also assume that the size of $V$ is small compared to $r$, which allows
us to expand everything in $1/r$.
For the metric tensor we obtain
\bq
\sqrt{|g|} & = & 1 + O\left(\frac{1}{r^2}\right)
\eq
and we consider therefore the action
\bq
S & = & \frac{1}{2 g^2} \int\limits_{V} d^4x  
\mbox{Tr} \; \hat{F}_{\mu\nu} \hat{F}^{\mu\nu}.
\eq
(For a treatment of gauge theories on curved manifolds without the above approximation we
refer to \cite{Adams}.)

\section{The action in a background field}

In this paragraph we express the Lagrangian
\bq
\label{Ltotal}
{\cal L}_{YM} & = & \frac{1}{2 g^2}  
\mbox{Tr} \; F_{\mu\nu}(A+B) F_{\mu\nu}(A+B),
\eq
corresponding to a gauge field configuration $A_\mu+B_\mu$ in terms
of a Lagrangian for the fluctuating field $A_\mu$ alone.
For the field strength $F_{\mu\nu}(A+B)$ we write
\bq
\label{fieldstrength}
F_{\mu\nu}(A+B) = F_{\mu\nu}(A) + K_{\mu\nu},
\eq
where
\bq
K_{\mu\nu} & = & \left[ A_\mu,B_\nu\right]-\left[A_\nu, B_\mu \right] .
\eq
We assumed that $B_\mu$ is a pure gauge field and therefore $F_{\mu\nu}(B)=0$.
Substitution of eq. (\ref{fieldstrength}) into eq. (\ref{Ltotal}) gives us
\bq
\label{L2}
{\cal L}_{YM} & = & \frac{1}{2 g^2} \mbox{Tr} \;
F_{\mu\nu}(A) F_{\mu\nu}(A) + 2 K_{\mu\nu} F_{\mu\nu}(A) + K_{\mu\nu} K_{\mu\nu},
\eq
We further assume that the instantons numbers of the configurations $A_\mu$ and $A_\mu+B_\mu$
are the same:
\bq
\label{instanton}
Q(A+B) & = & Q(A), 
\eq
where
\bq
Q(A) & = & \frac{1}{32 \pi^2} \eps_{\mu\nu\rho\sigma} \int d^4 x \mbox{Tr}\;
F_{\mu\nu}(A)  F_{\rho\sigma}(A).
\eq
We are primarily interested in configuration for which $Q(A+B)=Q(A)=0$. 
Clearly, for these configurations eq. (\ref{instanton}) is satisfied.
From eq. (\ref{instanton}) we obtain
\bq
\label{inst20}
- \frac{1}{4 g^2} \eps_{\mu\nu\rho\sigma} \int d^4 x \; \mbox{Tr} \;
2 K_{\mu\nu} F_{\rho\sigma}(A) + K_{\mu\nu} K_{\rho\sigma} & = & 0.
\eq
We add the l.h.s. of eq. (\ref{inst20}) 
to eq. (\ref{L2}) and obtain
\bq
\label{L3}
\int d^4x \; {\cal L}_{YM} & = & \frac{1}{2 g^2} \int d^4 x \; \mbox{Tr} \;
F_{\mu\nu}(A) F_{\mu\nu}(A) + 4 K_{\mu\nu}^- F_{\mu\nu}(A) + 2 K_{\mu\nu}^- K_{\mu\nu}^-,
\eq
where $K_{\mu\nu}^-$ is the anti-self-dual part of $K_{\mu\nu}$:
\bq
K_{\mu\nu}^- & = & \frac{1}{2} \left( K_{\mu\nu} - \frac{1}{2} \eps_{\mu\nu\rho\sigma}
K_{\rho\sigma} \right)
\eq
Since all additional terms involve
commutators, the abelian case of a $U(1)$ gauge potential is trivial.
We restrict ourselves therefore to the $SU(2)$ gauge potential.
The presence of the background field modifies the terms bilinear and trilinear
in the fluctuating field $A_\mu$.
The Lagrangian eq.(\ref{L3}) is invariant under the combined transformation
\bq
A_\mu' & = & U^{-1} A_\mu U + U^{-1} \partial_\mu U, \nonumber \\
B_\mu' & = & U^{-1} B_\mu U .
\eq
For the gauge transformations $U$ we restrict ourselves to transformations which are connected 
to the identity $U=1$.
In order to fix the gauge we choose the background field gauge \cite{BFG1,BFG2} and add
a gauge fixing term
\bq
\label{gaugefixing}
{\cal L}_{GF} & = & \frac{1}{2 g^2} \mbox{Tr} \; 2
\left( \partial_\mu A_\mu + \left[B_\mu, A_\mu \right] \right)
\left( \partial_\nu A_\nu + \left[B_\nu, A_\nu \right] \right)
\eq
to the Lagrangian.
Putting everything together we obtain for the terms bilinear in $A_\mu$:
\bq
\label{Lfinal}
\int d^4 x \; \left( {\cal L}_{YM}|_{bilinear} + {\cal L}_{GF} \right) & = &
\frac{1}{2 g^2} \int d^4 x \;
A_\mu^a \left( S^{ab}_{\mu\nu} + T^{ab}_{\mu\nu} \right) A_\nu^b, 
\eq
where
\bq
S^{ab}_{\mu\nu} & = & \left( 
- \delta^{ab} \Box 
+ 2 i f^{abc} B^c_\rho \partial_\rho
- f^{eac} f^{ebd} B^c_\rho B^d_\rho 
\right) \delta_{\mu\nu}, \nonumber \\
T^{ab}_{\mu\nu} & = & 
2 i f^{abc} \left( B_\mu^c \partial_\nu - B_\nu^c \partial_\mu
                     + \eps_{\mu\nu\rho\sigma} B^c_\rho \partial_\sigma \right) \nonumber \\
& & - f^{eac} f^{ebd} \left( B^c_\mu B^d_\nu - B^d_\mu B^c_\nu 
                     + \eps_{\mu\nu\rho\sigma} B^c_\rho B^d_\sigma \right).
\eq
$S^{ab}_{\mu\nu}$ is symmetric in $(\mu,\nu)$, whereas $T^{ab}_{\mu\nu}$ is self-dual.
Since we used eq. (\ref{instanton}), no anti-self-dual term appears.

\section{Phenomenology of the toy model}

We now come back to our toy model and use the explicit expressions
given by the equations (\ref{strucconst}) and (\ref{background2})
for the structure constants and the background field.
We consider the case in which the rotation matrix $\Lambda_{\mu\nu}$
is trivial:
\bq
\Lambda_{\mu\nu} & = & \delta_{\mu\nu}
\eq
In that case the explicit expression for the background field
reads
\bq
B^a_\mu & = & \frac{2 i m}{r} \delta_{a\mu}
\eq
In the high-energy limit $k^2 >> m^2/r^2$, where $k$ is the momentum of the gauge boson,
we may neglect all terms involving background fields and our toy model reduces to
the standard Yang-Mills theory with massless gauge bosons.
In the low-energy limit we may neglect the partial derivatives
and only the mass term survives in the quadratic part of the Lagrangian.
To be precise this limit is valid (with all the approximations we made)
in the region
\bq
\frac{1}{r^2} << k^2 << \frac{m^2}{r^2}.
\eq
In other words we are probing distances, which are
small compared to $r$, but large compared to $r/m$.
We write the mass term as
\bq
{\cal L}_{mass} & = & \frac{m^2}{2 g^2 r^2} A^a_\mu M^{ab}_{\mu\nu} A^b_\nu
\eq
where the matrix $M^{ab}_{\mu\nu}$ is given by
\bq
M^{ab}_{\mu\nu} & = & \eps_{eac} \eps_{ebd}
\left( \delta_{\mu\nu} \delta_{\rho \sigma} +\delta_{\mu\rho} \delta_{\nu\sigma}
-\delta_{\mu\sigma} \delta_{\nu\rho} + \eps_{\mu\nu\rho\sigma} \right)
\delta_{c\rho} \delta_{d\sigma}
\eq
The mass matrix can be diagonalized by changing the variables according to
\bq
\left( \begin{array}{c} W^1_{\mu} \\ W^2_\mu \\ W^3_\mu \end{array} \right)
& = & 
\left( \begin{array}{ccc}
\frac{1}{\sqrt{2}} \eta_{1\mu\nu} & -\frac{1}{\sqrt{2}} \eta_{2\mu\nu} & 0 \\
\frac{1}{\sqrt{6}} \eta_{1\mu\nu} & \frac{1}{\sqrt{6}} \eta_{2\mu\nu} & -\frac{2}{\sqrt{6}} \eta_{3\mu\nu} \\
\frac{1}{\sqrt{3}} \eta_{1\mu\nu} & \frac{1}{\sqrt{3}} \eta_{2\mu\nu} & \frac{1}{\sqrt{3}} \eta_{3\mu\nu} \\
\end{array}
\right)
\left( \begin{array}{c} A^1_\nu \\ A^2_{\nu} \\ A^3_\nu \end{array} \right),
\eq
where the 't Hooft symbols \cite{tHooft} $\eta_{a\mu\nu}$ are defined in the appendix.
We obtain for the mass term
\bq
\label{mass}
{\cal L}_{mass} & = & \frac{m^2}{2 g^2 r^2} \left( W^1_\mu W^1_\mu + W^2_\mu W^2_\mu + 4 \; W^3_\mu W^3_\mu \right).
\eq
We see that $W^1$ and $W^2$ are degenerate in mass and that the ratio of the masses
of $W^1$ and $W^3$ is independent of the winding number and is given by
\bq
\frac{m_{W^1}^2}{m_{W^3}^2} & = & \frac{1}{4}.
\eq
At this point a comment about Lorentz symmetry is in order: Altough neither our original manifold
$S^1 \times S^3$ nor the specific choice of the $SU(2)$-background field is invariant
under $SO(4)$-rotations, the effective Lagrangian of the high-energy limit as well
as the effective Lagrangian eq. (\ref{mass}) of the low-energy limit posseses a $SO(4)$-symmetry.
However this will be no longer true in the intermediate range
\bq
k^2 \approx \frac{m^2}{r^2},
\eq
where we probe 
distances approximately equal to $r/m$.
In that case the complete expression eq.(\ref{Lfinal}) has to be used.\\
\\
Summary and critics of the toy model: In the low-energy limit the toy model predicts the mass
ratios $m_{W_1} = m_{W_2} = m_{W_3} / 2$. Furthermore, the toy model does not possess
a $SO(4)$-symmetry in the intermediate range. These two facts are hardly compatible
with observations. (For a survey on possible Lorentz-violating effects in QED see \cite{Carroll}.)
If we may neglect electromagnetic interactions we expect the
weak gauge bosons to be degenerate in mass. 

\section{Improvement of the toy model}

We may think about the manifold $S^1 \times S^3$ as a collection of three-dimensional
slices $S^3$. Let us say that $x_1$, $x_2$ and $x_3$ are coordinates on $S^3$ and $x_0$
is the normal coordinate.
For each slice the winding number of $S^3 \rightarrow SU(2)$ is fixed. 
For a non-trivial winding number we obtained a non-zero background field
at $x_1=x_2=x_3=0$. 
Our toy model has the additional property that this background field
stays constant as we go along the normal coordinate $x_0$.
This is an unneccessary assumption. We may allow that the orientation of the background
field changes as we pass along $x_0$ and replace $B_\mu$ by
\bq
B_\mu' & = & \Sigma(x_0)^{-1} B_\mu \Sigma(x_0) + \Sigma^{-1}(x_0) \partial_\mu \Sigma(x_0),
\eq
where $\Sigma(x_0)$ does depend on $x_0$, but not on $x_1$, $x_2$ or $x_3$.
Let us assume that we are interested in proccesses with a scale $k^2$. This will probe
a distance 
\bq
\Delta x_0 & = & \frac{\lambda}{\sqrt{k^2}}
\eq
in the normal direction, where $\lambda$ is some number between $0$ and $1$.
Let us assume that $\Delta x_0$ is sufficiently large, such that the map
\bq 
\Sigma & : & \left[ 0, \Delta x_0 \right] \rightarrow SU(2) \nonumber \\
 & & x_0 \rightarrow \Sigma(x_0)
\eq
sweeps out effectively all points in $SU(2)$-space. 
In this case it is reasonable to replace the $x_0$-dependent 
background field $B_\mu'$ by an average over all $SU(2)$-rotated configurations.
To see this let us assume that $\Delta x_0$ is made out of $n$ intervalls of length $\Delta$,
in which $B_\mu$ stays constant along the $x_0$ direction, and $n$ transition intervalls
of (negligible) length $\delta$, in which the background field changes from one orientation
to another. 
This can always be achieved by gauge transformations (connected to the identity).
Symbolically we have
\bq
\label{spliteq}
\lefteqn{
\int\limits_{\Delta x_0} dx_0 {\cal L}(\Sigma(x_0)^{-1} B_\mu \Sigma(x_0) 
+ \Sigma(x_0)^{-1} \partial_\mu \Sigma(x_0) ) 
 = } & & \nonumber \\
& = & n \cdot \Delta \sum\limits_i
{\cal L}(\Sigma(x^i_0)^{-1} B_\mu \Sigma(x^i_0) ) \nonumber \\
& & +
\int\limits_{n \times \delta} dx_0 {\cal L}(\Sigma(x_0)^{-1} B_\mu \Sigma(x_0) 
+ \Sigma(x_0)^{-1} \partial_\mu \Sigma(x_0) ) ,
\eq
where the sum is over all plateaux in which $B_\mu'$ is constant and
$x_0^i$ labels a point inside plateau $i$.
Up to gauge transformations and reparametrization the second term on the r.h.s of
eq. (\ref{spliteq}) is just $(n \cdot \delta / \Delta x_0)$ times the original
integral.
Therefore we established that we may replace the original integral over $x_0$ by
an avergage over all $SU(2)$ configurations.
Technically we do the averaging as follows: We replace $B_\mu$ by
\bq
B_\mu' & = & U^{-1} B_\mu U,
\eq
where
\bq
U & = & 1 z_0 + i \sigma_1 z_1 + i \sigma_2 z_2 + i \sigma_3 z_3
\eq
and
\bq
z_0 = \cos \alpha, & &
z_1 = \sin \alpha \cos \beta, \nonumber \\
z_2 = \sin \alpha \sin \beta \cos \gamma, & & 
z_3 = \sin \alpha \sin \beta \sin \gamma. 
\eq
In components we have
\bq
{B^a_{\mu}}' & = & \frac{1}{2} \mbox{Tr} \; \left( U \sigma_a U^{-1} \sigma_b \right) B_\mu^b
\eq
We then integrate over $\alpha \in [0,\pi]$, $\beta \in [0,\pi]$ and $\gamma \in [0, 2 \pi]$
with the measure
\bq
d \; U & = & \sin^2 \alpha \sin \beta \; d\alpha \; d\beta \; d\gamma.
\eq
We then obtain for the Lagrangian
\bq
\label{Limproved}
\int d^4 x \; \left( {\cal L}_{YM} + {\cal L}_{GF} \right) & = &
\frac{1}{2 g^2} \int d^4 x \;
A_\mu^a \left( - \Box + 2 \frac{m^2}{r^2} \right) A_\mu^a \nonumber \\
& & + 2 i f^{abc} A_\mu^a A_\nu^b \partial_\mu A_\nu^c 
- \frac{1}{2} f^{eac} f^{ebd} A_\mu^a A_\mu^b A_\nu^c A_\nu^d
\eq
We observe that all terms linear in the background field have dropped out and 
that the $SU(2)$-gauge bosons have aquired a mass $\sqrt{2} m/r$.
We also have observed that the effect of adding eq.(\ref{inst20}) to the Lagrangian eq.(\ref{L2})
drops out after averaging. 
Therefore, in principle, we could weaken assumption 1 and/or eq.(\ref{instanton}).
Finally, we would like to remark that the effective Lagrangians eq.(\ref{Lfinal}), eq.(\ref{mass})
and eq.(\ref{Limproved}) are not gauge invariant. Gauge invariance is broken by our choice
of the gauge fixing term in eq.(\ref{gaugefixing}).

\section{Conclusions}

In this paper we investigated the effect of the conservation of topological winding numbers.
We derived the effective theory in the background of a pure gauge field with non-trivial
winding number.
We showed that commutator terms may give rise to mass terms.
Therefore a $U(1)$-theory stays massless, whereas a $SU(2)$-theory develops massive modes.
In the most naive case we obtained an effective theory, which singles out a specific
orientation in $SU(2)$-space. Averaging over all orientations we obtained a Lorentz-invariant
effective theory where all $SU(2)$-gauge bosons aquire a mass $\sqrt{2} m / r$.

\section{Acknowledgements}

Many thanks to E. Laenen for organizing a working group at NIKHEF on effective
theories, from which this work originated.

\begin{appendix}

\section{The 't Hooft symbols}

The 't Hooft symbols $\eta_{a\mu\nu}$ and $\bar{\eta}_{a\mu\nu}$ are defined as \cite{tHooft}
\bq
\eta_{a\mu\nu} = \bar{\eta}_{a\mu\nu} = \eps_{a\mu\nu}, & & a, \mu, \nu = 1,2,3 \nonumber \\
\eta_{a\mu\nu} = - \eta_{a\nu\mu}, & & 
\bar{\eta}_{a\mu\nu} = - \bar{\eta}_{a\nu\mu}, \nonumber \\
\eta_{a\mu 0} = -\delta_{a\mu}, & & 
\bar{\eta}_{a\mu 0} = \delta_{a\mu}.
\eq
(Our notation differs slightly from 't Hooft, since we label the space-time coordinates
by $0,1,2,3$, whereas 't Hooft uses $1,2,3,4$.)
Our sign conventions for the antisymmetric tensors are: $\eps_{123}=+1$, $\eps_{0123}=+1$.
The tensor $\eta_{a\mu\nu}$ is self-dual, whereas $\bar{\eta}_{a\mu\nu}$ is anti-self-dual:
\bq
\eta_{a\mu\nu} = \frac{1}{2} \eps_{\mu\nu\rho\sigma} \eta_{a\rho\sigma}, & &
\bar{\eta}_{a\mu\nu} = -\frac{1}{2} \eps_{\mu\nu\rho\sigma} \bar{\eta}_{a\rho\sigma}.
\eq
We have the following relations:
\bq
\eta_{a\mu\nu} \eta_{b\mu\nu} = 4 \delta_{ab}, & &
\bar{\eta}_{a\mu\nu} \bar{\eta}_{b\mu\nu} = 4 \delta_{ab}, \\
\eta_{a\mu\rho} \eta_{a\nu\rho} = 3 \delta_{\mu\nu}, & &
\bar{\eta}_{a\mu\rho} \bar{\eta}_{a\nu\rho} = 3 \delta_{\mu\nu}, 
\eq
\bq
\eta_{a\mu\nu} \eta_{a\rho\sigma} & = & \delta_{\mu\rho} \delta_{\nu\sigma}
-\delta_{\mu\sigma} \delta_{\nu\rho} + \eps_{\mu\nu\rho\sigma}, \nonumber \\
\bar{\eta}_{a\mu\nu} \bar{\eta}_{a\rho\sigma} & = & \delta_{\mu\rho} \delta_{\nu\sigma}
-\delta_{\mu\sigma} \delta_{\nu\rho} - \eps_{\mu\nu\rho\sigma}, \\
\eta_{a\mu\rho} \eta_{b\nu\rho} & = & \delta_{ab} \delta_{\mu\nu} + \eps_{abc} \eta_{c\mu\nu}, \nonumber \\
\bar{\eta}_{a\mu\rho} \bar{\eta}_{b\nu\rho} & = & \delta_{ab} \delta_{\mu\nu} 
+ \eps_{abc} \bar{\eta}_{c\mu\nu} ,
\eq
\bq
\eps_{\mu\nu\rho\tau} \eta_{a\sigma\tau} & = & 
\delta_{\mu\sigma} \eta_{a\nu\rho} + \delta_{\nu\sigma} \eta_{a\rho\mu} 
+\delta_{\rho\sigma} \eta_{a\mu\nu}, \nonumber \\
\eps_{\mu\nu\rho\tau} \bar{\eta}_{a\sigma\tau} & = & 
- \delta_{\mu\sigma} \bar{\eta}_{a\nu\rho} - \delta_{\nu\sigma} \bar{\eta}_{a\rho\mu} 
-\delta_{\rho\sigma} \bar{\eta}_{a\mu\nu},
\eq
\bq
\eps_{abc} \eta_{b\mu\nu} \eta_{c\rho\sigma} & = & 
\delta_{\mu\rho} \eta_{a\nu\sigma} - \delta_{\mu\sigma} \eta_{a\nu\rho}
-\delta_{\nu\rho} \eta_{a\mu\sigma} + \delta_{\nu\sigma} \eta_{a\mu\rho}, \nonumber \\
\eps_{abc} \bar{\eta}_{b\mu\nu} \bar{\eta}_{c\rho\sigma} & = & 
\delta_{\mu\rho} \bar{\eta}_{a\nu\sigma} - \delta_{\mu\sigma} \bar{\eta}_{a\nu\rho}
-\delta_{\nu\rho} \bar{\eta}_{a\mu\sigma} + \delta_{\nu\sigma} \bar{\eta}_{a\mu\rho}, 
\eq
\bq
\eta_{a\mu\nu} \bar{\eta}_{b\mu\nu} & = & 0, \nonumber \\
\eta_{a\mu\rho} \bar{\eta}_{b\nu\rho} - \eta_{a\nu\rho} \bar{\eta}_{b\mu\rho} & = & 0.
\eq

\end{appendix}

\end{document}